\documentclass[twocolumn,notitlepage,nofootinbib,nobibnotes,aps,prd,10pt]{revtex4-1}
\usepackage{amsmath,amssymb, fp}
\usepackage{xcolor}
\definecolor{linkcolor}{RGB}{150,50,60}
\usepackage[	pdftex,colorlinks=true,	
			pdfstartview=FitV,
			linkcolor= linkcolor,
			citecolor= linkcolor,
			urlcolor= linkcolor,
			linktoc=page,
			hyperindex=true,
			hyperfigures=true,
			breaklinks=true]
			{hyperref}
\usepackage{dblfloatfix}
\usepackage{balance}
\usepackage{lastpage}
\usepackage{mathtools}
\usepackage{epsfig,rotating,pifont}
\usepackage{graphicx}
\usepackage[caption=false]{subfig}
\usepackage{bbold}
\usepackage{dsfont}
\usepackage{ragged2e}
\usepackage{etoolbox}
\usepackage{changepage}   
\usepackage{pgfplots}
\usetikzlibrary{pgfplots.groupplots}
\pgfplotsset{compat=1.3}
\usepackage{tikz}
\usetikzlibrary{arrows,shapes,positioning}
\usetikzlibrary{decorations.markings,decorations.pathmorphing,decorations.pathreplacing,decorations.text}
\usetikzlibrary{arrows,calc,patterns,shapes.geometric}
\usepackage{fancybox}
\usepackage{verbatim}
\usepackage{titlesec}

\def\beq{\begin{eqnarray}}
\def\eeq{\end{eqnarray}}

\def\a{\alpha}

\def\eps{\epsilon}

\def\la{\langle }
\def\ra{\rangle }

\def\lb{\label}

\def\M{\mathcal{M}}
\def\A{\mathcal{A}}
\def\B{\mathcal{B}}
\def\O{\mathcal{O}}
\def\P{\mathcal{P}}
\def\DM{\partial\mathcal{M}}

\newcommand{\be}{\begin{equation}}
\newcommand{\ee}{\end{equation}}
\newcommand{\bea}{\begin{eqnarray}}
\newcommand{\eea}{\end{eqnarray}}
\newcommand{\bg}{\begin{gather}}

\newcommand{\bseq}{\begin{subequations}}
\newcommand{\eseq}{\end{subequations}}

\renewcommand{\ln}{\mathop{\rm ln}\nolimits}

\def\tr{\hbox{Tr}}

\def\be{\begin{eqnarray}}
\def\ee{\end{eqnarray}}
\def\lb{\label}

\definecolor{Green}{RGB}{147,162,153}
\definecolor{Green2}{RGB}{26,148,49}
\definecolor{BrownL}{RGB}{173,143,103}
\definecolor{Red}{RGB}{210,83,60}
\definecolor{BrownD}{RGB}{114,96,86}
\definecolor{GreyD}{RGB}{76,90,106}
\definecolor{GreyB}{RGB}{128,141,160}
\definecolor{Maroon}{RGB}{121,70,61}

\definecolor{Blue}{RGB}{148,184,210}
\definecolor{Blue2}{RGB}{108,144,170}
\definecolor{Blue3}{RGB}{42, 107, 172}

\newsavebox\foobox
\begin{document}

\title{Boundary-corner entanglement for free bosons}

\author{Cl\'ement Berthiere$^{1,2,}$\vspace{3pt}}

\email{clement.berthiere@pku.edu.cn}

\affiliation{
$^1$Department of Physics, Peking University, Beijing 100871, China\\
$^2$Institut Denis Poisson, Universit\'e de Tours---CNRS, 37200 Tours, France
}

\date{November 30, 2018}

\begin{abstract}\vspace{4pt}
\begin{center}\textbf{\abstractname}\end{center}\vspace{-5pt}

In quantum field theories defined on a spacetime with boundaries, the entanglement entropy exhibits subleading, boundary-induced corrections to the ubiquitous area law. At critical points described by conformal field theories (CFTs), and when the entangling surface intersects the physical boundary of the space, new universal terms appear in the entropy and encode valuable information about the boundary CFT. In $2+1$ dimensions, the universal subleading boundary term is logarithmic with coefficient $b(\theta)$ depending on the angle $\theta$ at which the entangling surface intersects the boundary, as well as on the boundary conditions (BCs). 
In this paper, we conduct a numerical study of $b(\theta)$ for free bosons on finite-size square lattices. We find a surprisingly accurate fit between our lattice results and the corresponding holographic function available in the literature. We also comment on the ratio $b''(\pi/2)/A_T$, where $A_T$ is the central charge in the near boundary expansion of the stress tensor, for which a holographic analysis suggests that it may be a universal quantity. Though we show evidence that this ratio is violated for the free boson with Dirichlet BCs, we conjecture its validity for free bosons evenly split between Dirichlet and Neumann BCs.

\end{abstract}

\maketitle

\makeatletter
\def\l@subsubsection#1#2{}
\makeatother

\section{Introduction}

First introduced in the early 1930s by von Neumann \cite{vonNeumann}, the entanglement entropy was 60 years later put forward in an attempt to explain black hole entropy \cite{Bombelli:1986rw,Srednicki:1993im} and has since emerged as a prominent tool in many different areas of theoretical physics. 
In quantum field theory (QFT), the entanglement entropy of a spatial region $A$ is defined as the von Neumann entropy of the reduced density matrix $\rho_A$ on $A$, \mbox{$S_A=-\tr_A(\rho_A\ln\rho_A)$}.
This entropy is UV-divergent due to short-range correlations for generic states in QFTs and thus needs to be regulated by a cut-off $\eps$. In continuum Lorentz-invariant theories defined on a $(d+1)-$dimensional spacetime without boundary, the general structure of UV divergence for a smooth entangling surface $\Sigma$ (the boundary of the region $A$) takes the following form: 
\be
S_A = \gamma\frac{A(\Sigma)}{\eps^{d-1}} + \frac{s_{d-3}}{\eps^{d-3}} + \cdots + 
\begin{cases}\displaystyle
\;s_{log}\ln\eps\,,  &\;\;\, d+1\;\text{even} \nonumber\\
\; \displaystyle s_0\,, &\;\;\, d+1\;\text{odd}\nonumber
\end{cases}\nonumber
\ee
where the leading term obeys the area law \cite{Srednicki:1993im,Eisert:2008ur}. The entanglement entropy is particularly useful to probe the structure of conformal field theories (CFTs). In even dimension, the relevant subleading term to the area law in $S_A$ is a logarithmic divergence whose coefficient encodes information about the central charges of the theory---charges that appear in the trace anomaly \cite{Deser:1993yx,Duff:1993wm,Callan:1994py,Calabrese:2004eu,Solodukhin:2008dh}. The logarithmic term in even dimensions is thus closely related to the trace anomaly. 
In an odd-dimensional spacetime without boundary, the conformal anomaly trivially vanishes since it is impossible to construct invariants of odd dimension from the Riemann curvature and its derivatives. For similar reasons, the logarithmic term in the entropy for odd-dimensional CFTs is absent and the subleading term that contains relevant information about the theory is finite. However, the story is quite different for CFTs in the presence of boundaries. Boundary conformal field theories (BCFTs) have been extensively studied \cite{Cardy:1989ir,Cardy:2004hm,McAvity:1993ue,McAvity:1995zd,Fursaev:2015wpa,Huang:2016rol,Herzog:2017xha}, and constructions of their gravitational duals have also been discussed \cite{Takayanagi:2011zk,Fujita:2011fp,Nozaki:2012qd,Astaneh:2017ghi,Miao:2017gyt,Chu:2017aab}. For BCFTs, new anomalies localized on the boundary appear together with new central charges \cite{Fursaev:2015wpa,Solodukhin:2015eca,Huang:2016rol,Herzog:2017xha}. In particular, the conformal anomaly of an odd-dimensional BCFT is a pure boundary term. For example, in three dimensions, the integrated conformal anomaly reads \cite{Graham:1999pm,Jensen:2015swa, Solodukhin:2015eca} 
\be
\A_3\equiv\int_{\M_3}\la T\ra = \frac{1}{384\pi}\int_{\DM_3}\left( -a\hat{R} + \tfrac{3}{2}q\,\tr\,\hat{k}^2\right),\quad \lb{ano3}
\ee
where $a$ and $q$ are boundary central charges, $\hat{R}$ is the Ricci curvature on the boundary and $\hat{k}_{\mu\nu}$ is the traceless part of the extrinsic curvature tensor of the boundary. In parallel, it has been shown that the entanglement entropy for BCFTs acquires new subleading boundary-induced terms, see for instance \cite{2006PhRvL..96j0603L, Fursaev:2006ng, 2009JPhA...42X4009A,Hertzberg:2010uv, Fursaev:2013mxa, Jensen:2013lxa,Herzog:2015ioa, Fursaev:2016inw,Berthiere:2016ott} and \cite{Takayanagi:2011zk,Fujita:2011fp,FarajiAstaneh:2017hqv,Chu:2017aab,Seminara:2017hhh,Seminara:2018pmr} for holographic treatments. For particular entanglement geometries, the subleading corrections to the area law are purely boundary-induced, and one may then extract from the entanglement entropy the boundary contributions that encode universal features of the field theory.

In three-dimensional BCFTs, if one wishes to compute the entanglement entropy of some spatial region, one can envisage three different situations: \textit{i}) The entangling surface is a smooth curve that does not intersect the boundary of the space. In that case, there is no logarithmic divergence in the entropy. \textit{ii}) The region $A$ possesses sharp corners \cite{Casini:2006hu,Casini:2008as,Casini:2009sr,Myers:2012vs,Stoudenmire:2014hja,2014JSMTE..06..009K,Bueno:2015rda, Bueno:2015xda,Bueno:2015qya,Sahoo:2015hma,Helmes:2016fcp}, i.e. the entangling surface is singular, but does not intersect the boundary. The entanglement entropy then exhibits a logarithmic divergence whose coefficient, the corner function $a(\theta)$, depends on the opening angle of the corner and is recognized as an effective measure of the degrees of freedom in the underlying CFT \cite{Casini:2006hu,2014JSMTE..06..009K,Bueno:2015rda,Bueno:2015xda,Bueno:2015qya}. \textit{iii}) The entangling surface is smooth but intersects the boundary of the space. In that case, a different logarithmic contribution appears in the entanglement entropy \cite{Fursaev:2016inw,Berthiere:2016ott} as the region $A$ now has corners adjacent to the boundary. This kind of corners being qualitatively different than the sharp corners discussed just above, we will dub them ``boundary-corners''. When the opening angle of the boundary-corner is equal to $\pi/2$, the logarithmic coefficient in the entropy is controlled by the charge $a$ \cite{Fursaev:2016inw} present in the anomaly $\A_3$.  
In this manuscript, we are interested in the angle dependance of the boundary-corner function for free massless scalar fields.

\section{Boundary-corner entanglement}
In this paper, we study numerically the boundary-corner contribution to the entanglement entropy in $2+1$ dimensions for free massless scalar fields. The general geometrical setup we consider is depicted in \mbox{\hyperref[FIG1]{Fig.\,1}}: the spacetime has a flat boundary $\DM$, and the entangling surface $\Sigma$ that separates the region $A$ from its complement intersects $\DM$ at $\mathcal{C}$ with angle $\theta$. This configuration, besides the area law, captures the subleading boundary-induced logarithmic divergence only as $\Sigma$ is smooth (no sharp corners). 
\begin{figure}[h]
\centering
\includegraphics[]{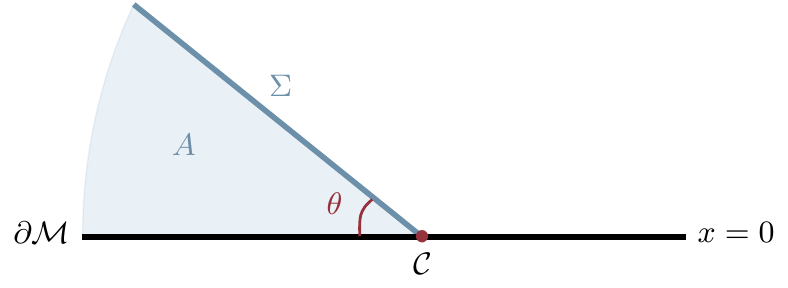}
\vspace{-5pt}
\caption{BCFT$_3$ on the half plane $x\ge0$ (time slice). The region $A$, shown in light blue, is bounded by $\Sigma$ (entangling surface, thick blue line) and $\partial\M$ (space boundary, thick black line). $\Sigma$ intersects $\partial\M$ at $\mathcal{C}$ with opening angle $\theta$.}
\lb{FIG1}
\end{figure}
For a region $A$ that \mbox{contains} multiple boundary-corners, i.e.\;for which $\Sigma$ intersects several times the boundary $\DM$, the coefficient of the logarithmic divergence is obtained by summing the contributions of all corners $\mathcal{C}_i$ on the boundary,  
\be
S_A = s_1\frac{L}{\epsilon} - \sum_{\mathcal{C}_i} b(\theta_i)\ln\frac{L}{\eps} + s_0 \,. 
\lb{slog3d}
\ee
Here $L$ is the length of $\Sigma$ and $\eps$ is a UV cutoff. For a conformally invariant theory, $s_{1,0}$ are non-universal constants, while $b(\theta)$ is universal and related to the trace anomaly \eqref{ano3}. 
Its form is constrained by properties of the entanglement entropy. Since we consider a BCFT$_3$ in its ground state which is a pure state, by symmetry $S_A=S_{\bar{A}}$ we have 
\be
b(\theta)=b(\pi-\theta)\,,\lb{pure}
\ee
which allows us to study this boundary-corner function for $0<\theta\le\pi/2$.
When the entangling surface is orthogonal to the boundary, the logarithmic term is given by \cite{Fursaev:2016inw} (see also \cite{Berthiere:2016ott} and \hyperref[apdxB]{Appendix} for scalar field results in $d$ spatial dimensions)
\be
b(\pi/2) = \frac{a}{24}\,,
\lb{slog3}
\ee
where $a$ is the boundary charge that appears in the conformal anomaly \eqref{ano3} and whose values may be found in \hyperref[tab1]{Table \ref{tab1}} for scalars and Dirac spinors. Assuming that the boundary-corner function is analytic about $\pi/2$, then it follows from the reflexion symmetry \eqref{pure} that $b(\theta)$ behaves around $\theta=\pi/2$ as:
\be
b(\theta\simeq\pi/2)= \frac{a}{24} + \sigma (\theta-\pi/2)^2 + \gamma(\theta-\pi/2)^4 +\cdots.\;\quad\lb{ortholim}
\ee
In the opposite limit $\theta\rightarrow 0$, $b(\theta)$ is singular as the partition of the system into two parts cannot be defined. We expect in this limit
\be
b(\theta\rightarrow 0) = \frac{\kappa}{\theta}+\cdots\,.\lb{cusp}
\ee
Both $\sigma$ and $\kappa$ are supposed to contain valuable information about the BCFT. This is believed to be true 1) because of the holographic result of \cite{Seminara:2017hhh} which shows that the second derivative at $\pi/2$ of the holographic boundary-corner function (i.e. the holographic version of $\sigma$) is related to the charge $A_T$ associated to the one point function of the stress tensor in the BCFT$_3$, and 2) by analogy with the corner function $a(\theta)$ for non-smooth entangling surface for which the leading coefficient in the smooth limit (analog to the limit \eqref{ortholim}) is universal and characterizes the number of degrees of freedom in the underlying CFT.

The objective of this paper is to compute numerically $b(\theta)$ for free bosons with Dirichlet boundary conditions for a certain range of opening angles. We then calculate the universal numbers $\sigma$ and $\kappa$ from our lattice computations. We eventually compare our exact results to the anomaly-derived formula of \cite{Fursaev:2016inw} and to the holographic boundary-corner function found in \cite{Seminara:2017hhh}. We present both aforementioned analytical formulas hereafter.

\medskip
\noindent\textbf{Boundary-corner function from the anomaly}\smallskip

As is well-known in two and four dimensions, the logarithmic term in the entanglement entropy can be exactly derived from the integrated anomaly using the replica method \cite{Ryu:2006ef}. However, in three dimensions the relation between the boundary anomaly $\A_3$ and the logarithmic contribution $b(\theta)$ appears to be less transparent. In \cite{Fursaev:2016inw}, Fursaev and Solodukhin showed that for scalar fields the entropy at $\theta=\pi/2$ derived from the integrated anomaly differs from that computed via the heat kernel. This mismatch is imputed to the occurrence of the non-minimal coupling of the scalar field to the curvature. It is further argue in \cite{Fursaev:2016inw} that the complete logarithmic term for arbitrary angle $\theta$ is the difference between the contribution obtained from the anomaly and the one coming from the non-minimal coupling. Their resulting proposed analytical formula for the boundary-corner function, which we denote $b_{FS}(\theta)$, is the following\footnote{We have corrected in \eqref{sloggen} a $\sin\theta$ factor in the $a$-part of $b_{FS}(\theta)$ that was missing in the calculations of \cite{Fursaev:2016inw}.}:
\be
b_{FS}(\theta) &=& \frac{a}{96\sin\theta}\left(3+\sin^2\theta\right) + \frac{q}{64}f(\theta)\,,\;
\lb{sloggen}
\ee
where $\displaystyle f(\theta)= \frac{1}{32}\frac{\cos^2\theta}{\sin\theta}\left(1+2\sin^2\theta+5\sin^4\theta \right)$ and $q$ is the second boundary central charge in $\A_3$. 
Formula \eqref{sloggen} possesses the qualitative behaviors expected for the boundary-corner function; it is a monotonic function of the opening angle from $\theta=0$ to $\theta=\pi/2$ (the decreasing/increasing character depends on the values of the charges $a$ and $q$), and near orthogonality one has
\be
b_{FS}(\theta\simeq\pi/2) = \frac{a}{24} + \frac{8a+3q}{768} (\theta-\pi/2)^2 + \cdots\,,\;
\ee
while in the cusp limit one obtains from \eqref{sloggen}
\be
b_{FS}(\theta\rightarrow 0) = \frac{64a+q}{2048\,\theta}+ \O(\theta)\,.\;
\ee
For a scalar field with Dirichlet boundary condition one gets
\be
\sigma^D_{FS}&=&\frac{11}{768}\simeq 0.01432\,,\\
\kappa^D_{FS}&=&\frac{65}{2048}\simeq 0.03174\,.
\ee

\begin{table}[h]\renewcommand{\arraystretch}{1.5}
\begin{center}
\vspace{0.2cm}
\begin{tabular}{|c|c|c|c|}
\hline
 Theory  & \;\;$a$\;\; & \;\;$q$\;\; & \,Boundary condition\,  \\ 
\hline \hline
 Real scalar & $1$ & $1$ & Dirichlet   \\
 \hline
 Real scalar & $-1$ & $1$ & Robin   \\
 \hline
Dirac spinor & $0$ & $2$ & mixed   \\
\hline
\end{tabular}
\vspace{-10pt}
\end{center}
\caption{Boundary central charges in the boundary conformal anomaly in 2+1 dimensions \cite{Solodukhin:2015eca,Fursaev:2016inw}.}
\label{tab1}
\end{table}

\medskip
\noindent\textbf{Holographic boundary-corner function}\smallskip

The holographic picture of AdS/BCFT was first introduced in \cite{Takayanagi:2011zk} and can be sketched as follows. The dual of a BCFT$_{d+1}$ is given by an asymptotically AdS$_{d+2}$ spacetime restricted by a $(d+1)$--dimensional brane $\mathcal{Q}$ whose boundary coincides with the boundary $\DM$ of the BCFT$_{d+1}$. Thus, the boundary of the bulk spacetime has two components, $\mathcal{Q}$ and the conformal boundary $\M$ on which lives the BCFT$_{d+1}$, and these two meet at a common boundary such that $\partial\mathcal{Q}=\DM$.
In this prescription \cite{Takayanagi:2011zk}, the gravitational action consists of the Einstein-Hilbert action to which is added a Gibbons-Hawking term and a boundary cosmological constant $T$ on the holographic boundary $\mathcal{Q}$. 

We consider a geometrical setup in which the boundary of the BCFT$_3$ is flat, and its extension $\mathcal{Q}$ into the bulk has constant tension $T$. The constant $T$ is related to the slope $\alpha$ of $\mathcal{Q}$ as $T = 2\cos \alpha$. 
Then as usual, according to Ryu-Takayanagi formula \cite{Ryu:2006bv}, the holographic entanglement entropy is proportional to the area of the minimal surface anchored on the entangling surface on the BCFT side---minimal surface which also ends on $\mathcal{Q}$. Hence, the entanglement entropy computed within the AdS/BCFT framework will in general depend on the slope $\alpha$ of $\mathcal{Q}$. In particular, the holographic boundary-corner function is not only a function of the opening angle $\theta$ but is also parametrized by $\alpha$.

In \cite{Seminara:2017hhh}, the authors computed the minimal surface corresponding to an infinite wedge with opening angle \mbox{$\theta \in\; ]0,\pi/2]$} having one of its edges on the boundary of the BCFT$_3$ and found an analytic expression for the holographic boundary-corner function, which they call $F_\alpha(\theta)$. The complete expression of $F_\alpha(\theta)$ is rather complicated; it involves elliptic integrals and is given in a parameterized form. Thus we will not reproduce their result here, and we refer the reader to the section 6 of \cite{Seminara:2017hhh} for details. Instead, we only report the limiting regimes of interest, i.e. the orthogonal and cusp limits:
\be
F_\alpha(\theta\simeq\pi/2)&=& -\cot\alpha + \frac{(\theta-\pi/2)^2}{2(\pi-\alpha)} + \cdots \,,\;\lb{Fortho}\\
F_\alpha(\theta\rightarrow0)&=& \frac{g(\alpha)^2}{\theta} + \O(\theta)\,,\lb{Fcusp}
\ee
where $g(\alpha)=E(\pi/4-\alpha/2 | 2) - \frac{\cos\alpha}{\sqrt{\sin\alpha}} + \frac{\Gamma(3/4)^2}{\sqrt{2\pi}}$, with $E(\phi | m)$ being the elliptic integral of the second kind.

In the following section, we introduce our calculation for the free massless scalar field theory on a finite-size square lattice. We will then compare our lattice results for the boundary-corner function to the analytical formulas proposed in \cite{Fursaev:2016inw} (field theoretic) and \cite{Seminara:2017hhh} (holographic).

\section{Free bosons on the lattice}
We consider the lattice Hamiltonian of a free massless scalar field given by
\be
H_d &=& \frac{1}{2}\sum_{\mathbf{x}}\Big[ \pi^2_{\mathbf{x}} + (\phi_{x_1+1,x_2,\cdots,x_d}-\phi_{\mathbf{x}})^2 + \cdots \nonumber\\
&& \hspace{2cm} + (\phi_{x_1,x_2,\cdots, x_d+1}-\phi_{\mathbf{x}})^2 \Big],
\lb{Hd}
\ee
where $d$ is the spatial dimension of the lattice, \mbox{$\mathbf{x} = (x_1, x_2, \cdots , x_d)$} represents the spatial lattice coordinates with $x_i=1, \cdots,L_i$, and $L_i$ is the lattice length along the $i^{th}$ direction. The total number of sites is $N=L_1L_2\cdots L_d$. The lattice spacing $\eps$ has been set to unity. The Hamiltonian \eqref{Hd} can be written as
\be
H_d = \frac{1}{2}\sum_{\mathbf{x}} \pi^2_{\mathbf{x}} +  \frac{1}{2}\sum_{\mathbf{x},\mathbf{y}}\phi_{\mathbf{x}}K_{\mathbf{x}\mathbf{y}}\phi_{\mathbf{y}} ,\lb{Hd2}
\ee
where $K$ is a positive-definite $N\times N$ matrix encoding the nearest-neighbor interactions between lattice sites as well as the boundary conditions. One is free to impose either e.g. periodic (PBCs) or open (OBCs) boundary conditions along each direction,
\begin{align}
{\rm PBCs:\quad} & \phi_{x_i} = \phi_{x_i+ L_i}  &{\rm and}& \quad \pi_{x_i} = \pi_{x_i+ L_i}\,,  \\
{\rm OBCs:\quad} & \phi_{0} = \phi_{L_i +1} = 0  &{\rm and}& \quad \pi_{0} = \pi_{L_i +1} = 0\,.
\end{align}
The vacuum two-point correlation functions are given by
\be
\la \phi_{\mathbf{x}}\phi_{\mathbf{x'}} \ra = \frac{1}{2}K^{-1/2}\,, \quad{\rm and}\quad \la \pi_{\mathbf{x}}\pi_{\mathbf{x'}} \ra  = \frac{1}{2}K^{1/2}\,. \lb{corr}
\ee
We are only interested in the elements of the correlation matrices for the distinguished region $A$ \cite{2003JPhA...36L.205P}:
\be
\left.\begin{aligned}
\;\left(X_A\right)_{ij} &= \la \phi_{\mathbf{x}_i}\phi_{\mathbf{x}_j} \ra \;\;\\
\;\left(P_A\right)_{ij} &= \la \pi_{\mathbf{x}_i}\pi_{\mathbf{x}_j} \ra\;\;
\end{aligned}\right\} \;\,\forall (\mathbf{x}_i,\mathbf{x}_j)\in A\,.
\ee
The entanglement entropy can then be calculated from the eigenvalues $\nu_\ell$ of the matrix $C_A=\sqrt{X_AP_A}$ \cite{Casini:2009sr}:
\be
S_d(A) &=& \sum_{\ell} \Bigg[\Big(\nu_\ell+\frac{1}{2}\Big)\log\Big(\nu_\ell+\frac{1}{2}\Big) \nonumber\\
&& \hspace{1.5cm} -\Big(\nu_\ell-\frac{1}{2}\Big)\log\Big(\nu_\ell-\frac{1}{2}\Big) \Bigg].\quad\lb{EE}
\ee
The procedure described above allows us to calculate the entropy for $two-$dimensional regions $A$ of sizes up to $\sim 4.10^4$ lattice sites on a computing server with 64GB RAM. In addition, for boundary-corners with opening angle $\theta=\pi/2$ (i.e.\;the entangling surface crosses the boundary orthogonally), we take advantage of the OBCs in one lattice direction to dimensionally reduce our $(2+1)-$dimensional model to multiple $(1+1)-$dimensional effective models. With this mapping, we can compute the entanglement entropy for much larger domains.
\bigskip

\noindent\textbf{Numerical extraction of boundary-corners}\smallskip

Let us discuss here the method that we employ to extract the numerical values of the boundary-corner function from the procedure outlined above.

The initial data are the values computed from \eqref{EE} for a range of $L$, where $L$ is the lattice length of the entangling surface. The final values for $b(\theta)$ are then obtained in two-steps, see e.g. \cite{Casini:2006hu,Casini:2009sr,Helmes:2016fcp,DeNobili:2016nmj}. First, we perform least-squares fits of these values to the general scaling ansatz,
\be
S(L)&=& s_1 L - b\ln L + s_0 \lb{fit}\\ 
&&\quad + s_{-1}L^{-1} + \cdots + s_{-p_{max}}L^{-p_{max}}\,,\nonumber
\ee
over various fit ranges $[L_{min},L_{max}]$ and $p_{max}\in[1,4]$, and obtain corresponding $b^{(L_{max})}(\theta)$ as functions of $L_{max}$. 
Second, we carry out an extrapolation of these values to the thermodynamic limit $L_{max}\rightarrow\infty$ and take the result as the final value for $b(\theta)$. 

As a warm-up, we first apply these methods in the following subsection in the case where the entangling surface crosses the boundaries orthogonally. The angle dependance is considered right after that.

\subsection{Orthogonal intersections}

\subsubsection{Dimensional reduction}\lb{dimred} 
\begin{figure}[h]
\centering
\subfloat[]{
\includegraphics[]{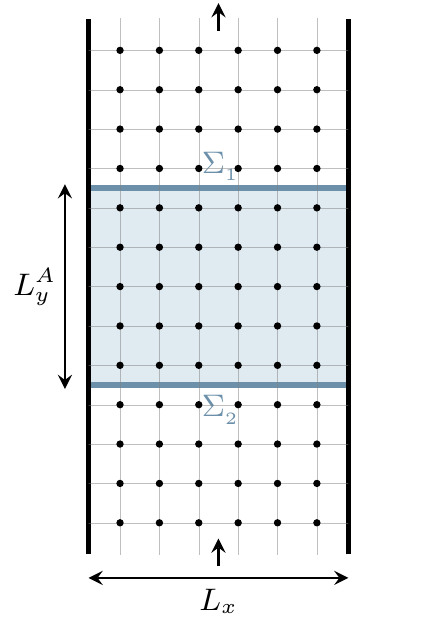}
\lb{FIG2a}}
\hspace{-0.7cm}
\subfloat[]{
\includegraphics[]{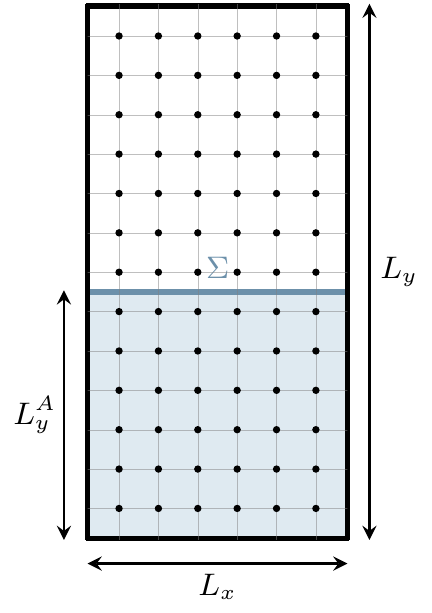}
 \lb{FIG2b}}
\caption{Two dimensional square harmonic lattices. The region $A$ is shown in light blue. OBCs are imposed in the horizontal $x$ direction in both cases. (a) PBCs is imposed in the vertical $y$ direction. The region $A$ is bounded by $\Sigma = \Sigma_1 \cup \Sigma_2$. (b) OBCs are imposed in the vertical $y$ direction. The region $A$ is bounded by $\Sigma$ (one component only).}
\end{figure}
In $d=2$ spatial dimensions, the region $A$ (see \mbox{\hyperref[FIG2a]{Fig.\,2}}) consists of $L_y^{A}$ complete rows, i.e.\;all sites along the $x$ direction for $L_y^{A}$ rows along $y$. One can decompose the fields $\phi_{x,y}$ and $\pi_{x,y}$ along the lattice direction $x$ such that
\be
\phi_{x,y} &=& \sqrt{\frac{2}{L_x+1}}\sum_{k_x} \sin(k_x x) \phi_y(k_x)\,,\\
\pi_{x,y} &=& \sqrt{\frac{2}{L_x+1}}\sum_{k_x} \sin(k_x x) \pi_y(k_x)\,,
\ee
where $\displaystyle k_x=\frac{n_x\pi}{L_x+1}$, with $n_x=1,\,\cdots,L_x$.
The Hamiltonian \eqref{Hd} in two spatial dimensions can then be
written as a sum over $L_x$ decoupled one dimensional Hamiltonians
\be
H_2 = \sum_{k_x}H_1(k_x)\,,
\ee
where the lower-dimensional Hamiltonians $H_1(k_x)$ are given by 
\be
H_1(k_x) &=& \frac{1}{2}\sum_{y}\bigg[ \pi_y^\dagger(k_x)\pi_y(k_x) + 4\sin^2\hspace{-2pt}\big(\tfrac{k_x}{2}\big)\phi_y^\dagger(k_x)\phi_y(k_x) \nonumber\\
&&\hspace{-0.7cm} + \Big(\phi_{y+1}^\dagger(k_x)-\phi_{y}^\dagger(k_x)\Big)\Big(\phi_{y+1}(k_x)-\phi_{y}(k_x)\Big) \bigg].
\ee
One notices that each Hamiltonian $H_1(k_x)$ corresponds to that of a one-dimensional free scalar field with effective mass $m^2_{k_x} = 4\sin^2(k_x/2)$. Therefore, the entropy $S_2(A)$ is given by a sum over $L_x$ different $k_x$--dependent entropies $S_1(L_y^A; k_x)$,
\be
S_2(A) = \sum_{k_{x}}S_1(L_y^A; k_x)\,.\lb{S1}
\ee
The highly time/memory consuming task of diagonalizing $N_A\times N_A$ matrices is thus reduced to diagonalizing $L_x$ matrices of size $L_y^A\times L_y^A$.
The generalization of the above procedure to higher dimensions is straightforward. We compute the boundary-corner function in $3+1$ dimensions in \hyperref[apdxC]{Section\;\ref{apdxC}}.

\subsubsection{Correlation functions} 
\paragraph{PBCs in the $y$ direction}
For the one-dimensional harmonic chain with effective mass $m^2_{k_{x}} = 4\sin^2(k_x/2)$ with PBCs in the $y$ direction, the two-point functions are given by
\begin{align}
(X_A)_{ij}\,\equiv\,\la \phi_i \phi_{j}\ra &= \frac{1}{2L_y}\sum_{k_y}\frac{1}{\omega_{\mathbf{k}}}\cos\big(k_y(i-j)\big)\,,\lb{XA}\\
(P_A)_{ij}\,\equiv\,\la \pi_i\pi_{j}\ra &= \frac{1}{2L_y}\sum_{k_y}\omega_{\mathbf{k}}\cos\big(k_y(i-j)\big)\,,\lb{PA}
\end{align}
where $\displaystyle \omega_{\mathbf{k}}^2= m^2_{k_x} +4\sin^2(k_y/2)$ and $\displaystyle k_y=2\pi n_y/L_y$ with $n_y=1,\,\cdots,L_y$ and $i,j \in A$. In the thermodynamic limit $L_y \rightarrow \infty$ \big(i.e. $\frac{1}{L_y}\sum_{k_y}\rightarrow \frac{1}{2\pi}\int_0^{2\pi}dk_y \big)$, these vacuum state correlators become \cite{2004PhRvA..70e2329B}
\be
(X_A)_{ij} &=& \frac{z^{i-j}}{2}\sqrt{\frac{z}{1-z^2}} \binom{i-j-1/2}{i-j}\lb{XAPBC}\\
&& \qquad\times \,{_2}F_1\bigg(\frac{1}{2},\frac{1}{2};i-j+1; \frac{z^2}{z^2-1}\bigg)\,,\nonumber\\
(P_A)_{ij} &=& \frac{z^{i-j}}{2}\sqrt{\frac{1-z^2}{z}}\binom{i-j-3/2}{i-j}\lb{PAPBC}\\
&& \qquad\times \,{_2}F_1\bigg(\hspace{-4pt}-\frac{1}{2},\frac{3}{2};i-j+1; \frac{z^2}{z^2-1}\bigg)\,,\quad\nonumber
\ee
where $\displaystyle z = \frac{1}{4}\Big(m_{k_{x}} - \sqrt{m_{k_{x}}^2+4}\Big)^2$.

\paragraph{OBCs in the $y$ direction}
If one imposes OBCs in the $y$ direction, the two-point functions are now given by
\begin{align}
(X_A)_{ij} &= \frac{1}{L_y+1}\sum_{k_y}\frac{1}{\omega_{\mathbf{k}}}\sin(k_y i)\sin(k_y j)\,,\lb{XAO}\\
(P_A)_{ij} &= \frac{1}{L_y+1}\sum_{k_y}\omega_{\mathbf{k}}\sin(k_y i)\sin(k_y j)\,,\lb{PAO}
\end{align}
where $\displaystyle \omega_{\mathbf{k}}^2= m^2_{k_{x}} +4\sin^2(k_y/2)$ and $\displaystyle k_y=n_y\pi/(L_y+1)$ with $n_y=1,\,\cdots,L_y$ and $i,j \in A$.
Due to the presence of the non-zero effective mass $m_{k_x}$, we have not been able to find analytical expressions for these two correlators in the thermodynamical limit.

Summarizing, the numerical evaluation of the entropy for a massless scalar field for the spatial configurations depicted \mbox{\hyperref[FIG2a]{Fig.\,2}} starts with the calculation of the $L_y^A\times L_y^A$ correlation matrix $C_A$ from (\ref{XA}-\ref{PA}) or (\ref{XAO}-\ref{PAO}) for a given effective mass $m_{k_x}$. Then, we calculate the contribution $S_1(L_y^A; k_x)$ in \eqref{S1}, and finally, the entropy is given by the sum \eqref{S1}.

\subsubsection{Results}

We have calculated the entanglement entropy for regions orthogonal to the boundaries of sizes up to $60\times 1000$ ($L_y^A=1000$). The numerical results are shown in \mbox{\hyperref[figortho]{Fig.\,3}} where we have plotted the logarithmic contribution found from our fitting procedure against $L_x^{max}$. We obtain from our best fits and extrapolations:
\be
y{\rm-PBCs:}\quad\; b &=& 0.16666(6) \simeq \frac{4}{24} = 4 b(\pi/2)\,,\quad \\
y{\rm-OBCs:}\quad\; b &=& 0.083333(3) \simeq \frac{2}{24} = 2 b(\pi/2)\,.\;\quad
\ee
We thus find a perfect agreement to the sixth digit between the lattice numerics and the field theory results $b(\pi/2)=a/24$ with $a=1$ for the free boson with Dirichlet BCs (OBCs). The factor 4 (2) comes from the fact that the entangling surface intersects orthogonally the boundaries 4 (2) times in the case where we impose PBCs (OBCs) in the $y$ direction (see \mbox{\hyperref[FIG2a]{Fig.\,2}}). The total logarithmic contribution in the entropy is the sum of all the intersections, hence the factors 4 or 2.
We consider next the angle dependence. 

\begin{figure}[h]
\centering
\includegraphics[scale=0.995]{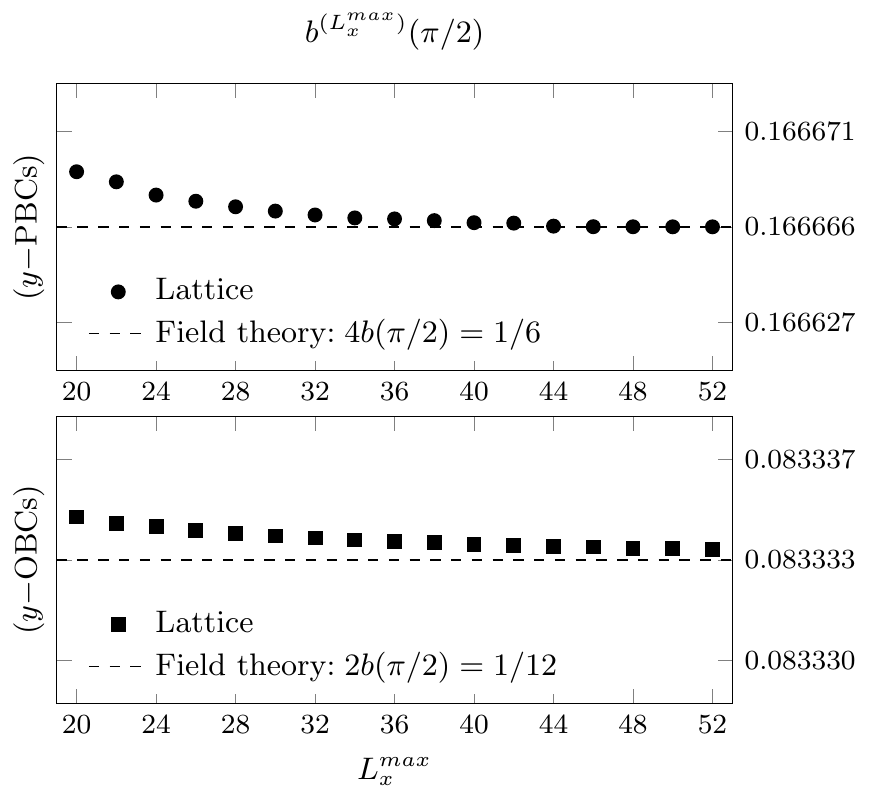}
\vspace{-18pt}
\caption{Convergence analysis of the fitted values of the boundary-corner contributions as a function of $L_x^{max}$ ($p_{max}=3$). The configurations employed here are shown \mbox{\hyperref[FIG2a]{Fig.\,2}}. Top: PBCs are imposed in the $y-$direction. Bottom: OBCs are imposed in the $y-$direction. The horizontal dashed lines correspond to the expected values from field theory \eqref{slog3} with $a=1$.} 
\lb{figortho}
\end{figure}

\subsection{Angle dependence}

When the entangling surface is not orthogonal to the boundaries, we cannot sine-decompose the fields in the $x$-direction as we did in the previous section and hence dimensionally reduce our numerical problem. 
Nevertheless, we may still compute the entanglement entropy on full two-dimensional lattices for configurations shown in \mbox{\hyperref[FIG3a]{Fig.\,\ref{FIG3a}}}. In the $y$--direction we impose PBCs as this allows us to access the thermodynamic limit $L_y\rightarrow\infty$ \mbox{analytically}. For OBCs and PBCs in the $x$-direction and $y$-direction, respectively, the two-dimensional vacuum two-point functions in the thermodynamic limit $L_y\rightarrow\infty$ are the following:
\be
\la \phi_{i,j} \phi_{r,s}\ra  &=&  \frac{\binom{j-s-1/2}{j-s}}{L_x+1}\sum_{k_x}\sin(k_x i)\sin(k_x r)\lb{Xang}\\
&& \hspace{-10pt}\times\sqrt{\frac{z^{2(j-s)+1}}{1-z^2}} \,{_2}F_1\bigg(\frac{1}{2},\frac{1}{2};j-s+1; \frac{z^2}{z^2-1}\bigg)\,,\nonumber\\
\la \pi_{i,j}\pi_{r,s}\ra &=&  \frac{\binom{j-s-3/2}{j-s}}{L_x+1}\sum_{k_x}\sin(k_x i)\sin(k_x r)\lb{Pang}\\
&& \hspace{-10pt}\times \sqrt{\frac{1-z^2}{z^{2(s-j)+1}}}\,{_2}F_1\bigg(\hspace{-4pt}-\frac{1}{2},\frac{3}{2};j-s+1; \frac{z^2}{z^2-1}\bigg)\,,\nonumber
\ee
where $\displaystyle k_x=\pi n_x/(L_x+1)$ with $n_x=1,\,\cdots,L_x$, and we defined $\displaystyle z \equiv z(k_x) = \Big(|\sin(k_x/2)| - \sqrt{\sin^2(k_x/2)+1}\Big)^2$.
Expressions \eqref{Xang} and \eqref{Pang} are the matrix elements of the correlation matrices $X_A$ and $P_A$ respectively (where $(i, j)$ and $(r, s)$ are the raw and column indices respectively). The entanglement entropy is calculated with \eqref{EE}.

We compute the entanglement entropy of regions $A$ of width $L_y^A\gg 1$ and length $L_x\ll L_y^A$ on a lattice with OBCs in the $x$-direction and PBCs in the $y$-direction (see \mbox{\hyperref[FIG3a]{Fig.\,\ref{FIG3a}}}). For such configurations, there are four identical boundary-corners due to the symmetry $b(\theta)=b(\pi-\theta)$. Then one extracts the logarithmic contribution in the entropy and divides it by four to get $b(\theta)$.
\begin{figure}[h]
\centering
\subfloat[]{
\includegraphics[]{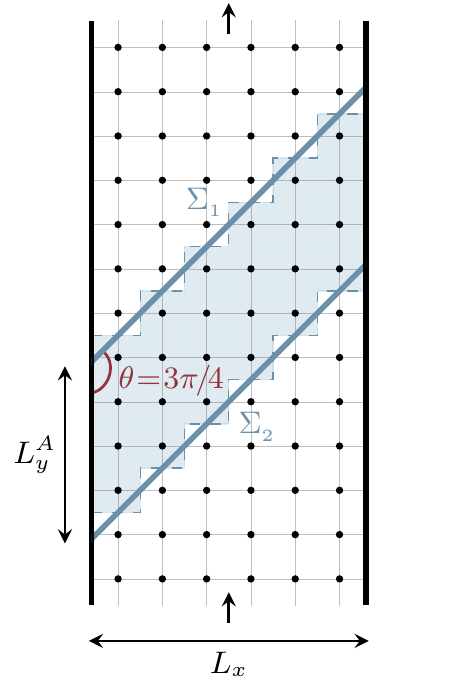}
}
\hspace{-0.8cm}
\subfloat[]{
\includegraphics[]{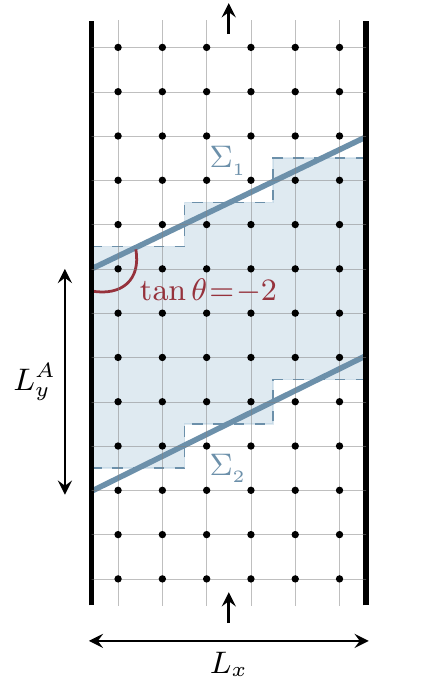}
}
\caption{Two-dimensional square lattices with OBCs imposed in the horizontal $x$ direction and PBCs in the vertical $y$ direction. The regions $A$ (blue) are bounded by $\Sigma = \Sigma_1 \cup \Sigma_2$. \mbox{(a) The entangling} surface intersects the boundaries with angles $\theta=3\pi/4$ and $\pi/4$. (b) The entangling surface intersects the boundaries with angles $\theta=\arctan(\pm 2)$.}
\lb{FIG3a}
\end{figure}
To access the boundary-corner coefficient of angles other than $\theta=\pi/2$, we follow e.g. \cite{Helmes:2016fcp} and \mbox{``pixelate"} the entangling surface whose intersections with the boundaries define the boundary-corners. Angles which obey $\tan\theta = r\in \mathbb{Q}$ are then accessible on square lattices. This is shown in \mbox{\hyperref[FIG3a]{Fig.\,\ref{FIG3a}}} for $\theta=3\pi/4$ (or $\pi/4$, equivalently) and $\tan\theta=\pm2$. For angles such that $\tan\theta=r>1$, the data points with which we perform our fits are chosen to be distant from each others by $\Delta L=r$. The errors related to the finite size of the lattices or those introduced through our choice of fitting procedures make the confidence intervals of our numerical values difficult to determine. The lattice results for the free boson with Dirichlet BCs are given in \hyperref[tab2]{Table \ref{tab2}}. We have reported there the digits that we found to be stable.
From our numerical data points, we obtain the following values for the orthogonal and cusp limit coefficients
\be
\sigma^D = 0.023(4)\,,\qquad \kappa^D = 0.044(4)\,.\quad\lb{sigma}
\ee
We have extracted the value $\sigma^D=0.023(4)$ from our lattice calculations by fitting our data to \eqref{ortholim} for $\tan\theta=4,5,6,7,8,\infty$. Similarly, we obtained $\kappa^D=0.044(4)$ by fitting our data points to \eqref{cusp} for $\tan\theta=1/8,1/6,1/4$.
\begin{table}[h]\renewcommand{\arraystretch}{1.5}
\begin{center}
\vspace{0.2cm}
\begin{tabular}{|c|c|c||c|c|c|}
\hline
 \;$\tan\theta$\; & \,\;lattice\,\; & \;\;FS\;\; & \,$\tan\theta$\, & \,\,\;lattice\,\,\; & \;\;FS\;\;  \\ 
\hline \hline
1/8 & $\,0.357(7)\,$ & \;$0.25723$\; & $3$ & $0.04419$ & $0.043175$  \\
\hline
1/6 & $0.269(7)$ & $0.19485$ & $4$ & $0.04310$ & $0.042533$  \\
\hline
1/4 & $0.182(4)$ & $0.13352$ & $5$ & $0.0425(9)$ & $0.042227$  \\
\hline
1/3 & $0.1395$ & $0.10385$ & $6$ & $0.0423(1)$ & $0.042058$  \\
\hline
1/2 & $0.09798$ & $0.075933$ & $7$ & $0.0421(4)$ & $0.041955$   \\
\hline
1 & $0.06080(9)$ & $0.052682$ & $8$ & $0.0420(3)$ & $0.041888$  \\
\hline
$2$ & $0.04717$ & \;$0.044888$\; & $\infty$ & \;$0.041666$\; & \,$0.0416666$\,  \\
\hline
\end{tabular}
\vspace{-10pt}
\end{center}
\caption{Lattice results for the boundary-corner function $b(\theta)$ for the free boson with Dirichlet BCs. Also indicated are values from FS function \eqref{sloggen}.}
\label{tab2}
\end{table}

\medskip
\noindent\textbf{Comparison with the literature}\smallskip

Comparing our numerical results with FS function \eqref{sloggen} is straightforward, see \hyperref[tab2]{Table \ref{tab2}} and \mbox{\hyperref[FIG5]{Fig.\,\ref{FIG5}}} where we have plotted our lattice results and $b_{FS}(\theta)$. We find a relatively good agreement between the two for angles $\theta>\pi/3$, for which the deviation from the numerics falls below $5\%$. However, in the small angle regime, the difference between the lattice data and the FS formula goes up to $30\%$, as one can see from the difference between $\kappa^D=0.044(4)$ and $\kappa^D_{FS}=65/2048\simeq0.03174$. If we compare $\sigma^D=0.023(4)$ and $\sigma^D_{FS}=11/768\simeq0.01432$, the latter differs by nearly $40\%$ with respect to the former. Overall, $b_{FS}(\theta)$ does not appear to describe consistently our lattice results over the whole range $0<\theta\le\pi/2$.

Let us now turn ourselves to the analytical formula derived within the AdS/BCFT framework in  \cite{Seminara:2017hhh}. To compare their holographic boundary-corner contribution $F_\alpha(\theta)$ with our numerical free boson result $b(\theta)$, we must start by normalizing $F_\alpha(\theta)$. A convenient choice is to consider the function $b_\a(\theta)$ defined as 
\be
\frac{b_{\alpha}(\theta)}{b(\pi/2)} \equiv \frac{F_\alpha(\theta)}{F_\alpha(\pi/2)}\,, \lb{bholo}
\ee
such that $b_{\alpha}(\pi/2)=b(\pi/2)$. The extra parameter $\a$ is related to the boundary conditions in the holographic field theory, but a precise dictionary between them has yet to be established. The holographic BCFT is a strongly coupled theory with a large number of degrees of freedom. In contrast, the free boson has only one degree of freedom and zero coupling. Furthermore the BCs of the holographic theory are certainly more complicated than the Dirichlet one we consider for the free boson. It may thus seem unlikely that holographic results reproduce exactly free field ones given how different the two field theories are.
Comparison of $b_\alpha(\theta)$ to our lattice data shows a surprising excellent match between the two for a specific value of $\alpha$, namely $\alpha_s\simeq2.56(3)\simeq 147^\circ$. As one can see in \hyperref[FIG5]{Fig.\,\ref{FIG5}}, the resulting normalized holographic function $b_{\a_s}(\theta)$ and our data for the free boson with Dirichlet BCs agree with each other exceptionally well, within less than 0.1$\%$ discrepancy, for the range of angles we considered. 
We can also use the liming regimes \eqref{Fortho} and \eqref{Fcusp} of the holographic boundary-corner function to get estimates of $\sigma^D$ and $\kappa^D$. For $\a=\a_s$, $b_{\a_s}(\theta)$ yields the values $\sigma^D=0.023(47)$ and $\kappa^D=0.044(35)$ which are indeed very close to the ones we obtained in \eqref{sigma}. 
\begin{figure}[h]
\centering\hspace*{-7pt}
\includegraphics[scale=1]{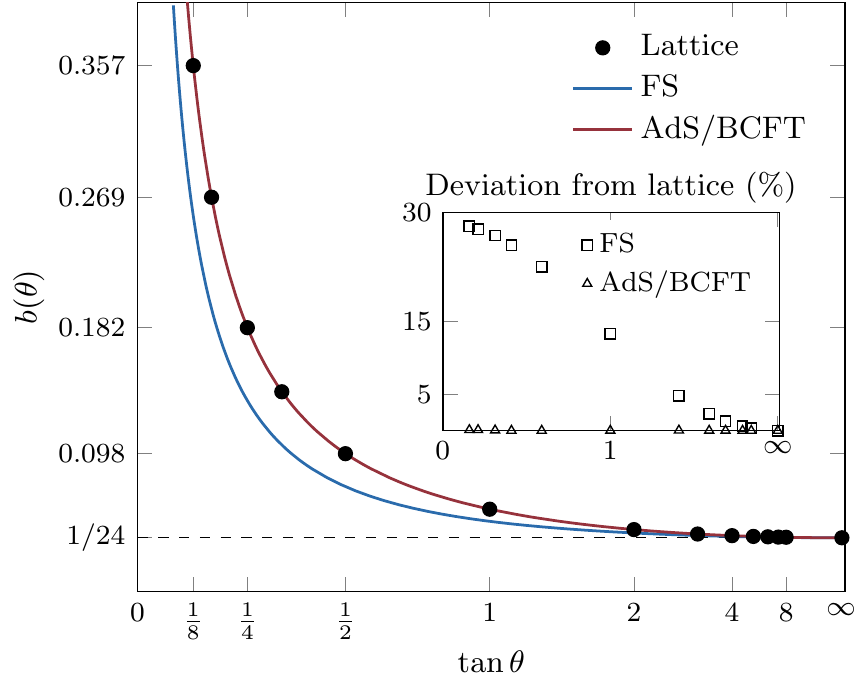}
\vspace{-11pt}
\caption{Boundary-corner entanglement for the free boson with Dirichlet BCs. Our lattice results (black dots) are compared to the (normalized) holographic function $b_{\a}$ with \mbox{$\a=2.56(3)$} (red) and FS function $b_{FS}$ (blue). The inset shows the deviation of the two functions from each data point.} 
\lb{FIG5}
\end{figure}

\medskip

\noindent\textbf{A universal ratio? Holography vs. free field}\smallskip

In a BCFT$_3$, the near boundary behavior of the one point function of the stress tensor is given by \cite{Deutsch:1978sc}
\be
\la T_{ij} \ra = \frac{A_T}{x^2}\hat{k}_{ij} + \cdots\,,\qquad x\rightarrow0\,,
\ee
where $x$ is the proper distance from the boundary, $\hat{k}_{ij}$ is the traceless part of the extrinsic curvature of the boundary and the coefficient $A_T$ \textit{a priori} depends on the boundary conditions of the BCFT$_3$. In \cite{Seminara:2017hhh}, it has been observed that for holographic theories dual to Einstein gravity, the ratio $\frac{1}{4G_N}F''_\alpha(\pi/2)/A_{T,E}$ is independent of $\alpha$,
\be
\sigma_{E}/A_{T,E}=-\pi\,, \lb{Hratio}
\ee
where we have introduced $\sigma_E=\frac{1}{8G_N}F''_\alpha(\pi/2)$ and the subscript $E$ means that this quantity was computed holographically for a bulk theory described by Einstein gravity. As noted in \cite{Seminara:2017hhh} and above, it is indeed interesting that the ratio \eqref{Hratio} is independent of the slope $\alpha$ of the boundary because $\alpha$ is supposed to encode the boundary conditions of the dual BCFT in the AdS/BCFT picture. On the field theory side, $A_T$ has been computed in \cite{Miao:2018dvm} for free scalar fields with Dirichlet and Robin BCs, giving the same value $A_T = -1/128\pi$ for both BCs.
We thus find from our lattice calculations
\be
\sigma^D/A_T = -9.4 \simeq -3\pi\,, \lb{Fratio}
\ee
which is approximately three times the holographic ratio. Though one has to be careful when evaluating $\sigma^D$ from our data, we are confident that our lattice values of $b(\theta)$ are accurate up to their third/fourth significant digits, which does not allow for the value of $\sigma^D$ needed to satisfy \eqref{Hratio}.
It is somehow surprising that the holographic and free field ratios do not match. Indeed, one may have anticipated a similar outcome to that of the corner function $a(\theta)$ (see again \cite{2014JSMTE..06..009K, Bueno:2015rda, Bueno:2015xda}) for which holography and field theory agree for the value of the ratio between the leading coefficient in the smooth limit and the central charge $C_T$ of the CFT. 
However, one may suggest\footnote{I thank R.-X. Miao for discussions on this point.} that for $\alpha=\pi/2$ the holographic BCFT$_3$ shares some common properties with free scalars, half of them with Dirichlet BCs and the other half with Robin BCs (e.g. same structures of one and two point functions \cite{McAvity:1993ue,Alishahiha:2011rg,Seminara:2017hhh,Miao:2018dvm} and vanishing logarithmic contribution at orthogonality in the holographic entanglement entropy \cite{FarajiAstaneh:2017hqv, Seminara:2017hhh}). One may therefore conjecture that the holographic ratio \eqref{Hratio} could still be valid for free bosons with half Dirichlet and half Robin BCs. In that case, free scalars with Robin BCs would have to satisfy the ratio $(\sigma^D+\sigma^R)/2A_T = -\pi$, from which one may estimate the Robin coefficient to be $\sigma^R\simeq-0.0078$.
It would be very interesting to test this conjecture for the Robin case with field theoretic calculations.

\section{Boundary-corners in 3+1 dimensions}
\begin{figure}[h]
\begin{center}
\includegraphics[scale=1]{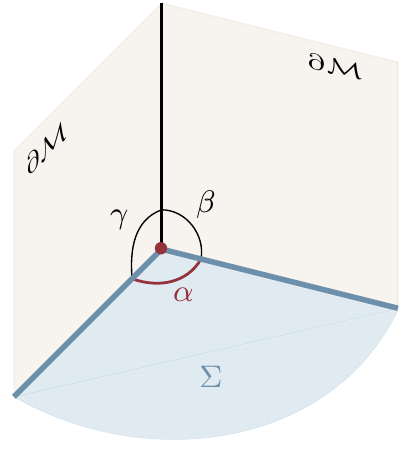}
\end{center}
\vspace{-20pt}
\caption{$3d$ boundary-corner. The intersection of the entangling surface $\Sigma$ (blue) with the boundary $\partial\cal{M}$ is a broken line (thick blue). The boundary-corner is marked by a red dot.}
\label{fig2Ap}
\end{figure}

\lb{apdxC}
Boundary-corners are not exclusive to (2+1) dimensional spacetimes but can also be defined in higher dimensions. In three spatial dimensions, a boundary-corner is the intersection of three two-dimensional surfaces where two of them are boundaries of the space and the other one is the entangling surface. Such a corner is shown in  \mbox{\hyperref[fig2Ap]{Fig.\,\ref{fig2Ap}}}. In general, the corresponding corner function $b_{3d}(\a,\beta,\gamma)$ multiplying the logarithmic divergence in the entanglement entropy depends on three opening angles $\a,\beta,\gamma$, as well as on the BCs.
Restricting ourselves to the case where the entangling surface is orthogonal to the boundaries, that is we fix $\beta=\gamma=\pi/2$ but $\alpha$ remains arbitrary, the boundary-corner function reduces to \cite{Kac, Hertzberg:2010uv}
\be
b_{3d}(\a,\pi/2,\pi/2) \equiv b_{3d}(\a) = \frac{\pi^2-\a^2}{144\pi\a}\,,\quad\lb{slog3a}
\ee
for both Neumann and Dirichlet BCs.
From the field theory results of \mbox{\hyperref[apdxB]{Appendix}}, the entanglement entropy for an orthogonal bipartition of the space with infinite square-cylindrical boundary in $3+1$ dimensions (see \mbox{\hyperref[fig1Ap]{Fig.\,\ref{fig1Ap}}}) reads 
\be
S_3(L) = s_2\frac{L^2}{\eps^2}+s_1\frac{4L}{\eps}+ 4b_{3d}(\pi/2) \ln\frac{L}{\eps}+s_0\,,\quad\lb{S3}
\ee
where $b_{3d}(\pi/2)=1/96$ for a free massless scalar field. We would like to compute numerically $b_{3d}(\pi/2)$ for free bosons with Dirichlet BCs on the lattice. To do so, let us first generalize the procedure outlined in \hyperref[dimred]{subsection\,\ref{dimred}} to $d$ dimensions. 

The fields $\phi_{\mathbf{x}}$ and $\pi_{\mathbf{x}}$ are decomposed along the lattice directions $\mathbf{x}_d=(x_1,\cdots,x_{d-1})$ such that
\be
\phi_{\mathbf{x}} &=& \frac{2^{(d-1)/2}}{\sqrt{(L_{1}+1)\cdots(L_{d-1}+1)}}\\
&& \times \sum_{\mathbf{k}_{d-1}} \sin(k_{1} x_1)\cdots\sin(k_{d-1}x_{d-1}) \,\phi_y(\mathbf{k}_{d-1})\,,\nonumber\\
\pi_{\mathbf{x}} &=& \frac{2^{(d-1)/2}}{\sqrt{(L_{1}+1)\cdots(L_{d-1}+1)}}\\
&& \times \sum_{\mathbf{k}_{d-1}} \sin(k_{1} x_1)\cdots\sin(k_{d-1}x_{d-1})\, \pi_y(\mathbf{k}_{d-1})\,,\nonumber
\ee
where $\mathbf{k}_{d-1}=(k_1,\,\cdots,k_{d-1})$ and $\displaystyle k_{i}=n_i \pi/(L_{i}+1)$, with \mbox{$n_{i}=1,\,\cdots,L_{i}$}.
The Hamiltonian \eqref{Hd} in $d$ spatial dimensions can be expressed as a sum over decoupled Hamiltonians in one spatial dimension,
\be
H_d = \sum_{\mathbf{k}_{d-1}}H_1(\mathbf{k}_{d-1})\,,
\ee
where each lower-dimensional Hamiltonian $H_1(\mathbf{k}_{d-1})$ corresponds to that of a one-dimensional free scalar field with mass \mbox{$m^2_{\mathbf{k}_{d-1}} = 4\sin^2(k_{1}/2)+\cdots+4\sin^2(k_{d-1}/2)$}. The entropy $S_d(A)$ is thus given by the multiple sum
\be
S_d(A) = \sum_{\mathbf{k}_{d-1}}S_1(L_y^A; \mathbf{k}_{d-1})\,.
\ee
In $d=3$ spatial dimensions with coordinates $(x_1, x_2, y)$, the entanglement entropy is given by a double sum over $k_1$ and $k_2$. We choose to set PBCs along $y-$direction and consider the thermodynamic limit $L_y\rightarrow\infty$ so that the $1d$ correlations matrices are given by \eqref{XAPBC} and \eqref{PAPBC} with \mbox{$m^2_{\mathbf{k}_2}=4\sin^2(k_{1}/2)+4\sin^2(k_{2}/2)$}. We set the lengths of the entangling surface in both directions $x_1$ and $x_2$ to be the same, i.e. $L_1=L_2=L$. Then for finite $L_y^A$ the entropy is twice that of \eqref{S3} because the entangling surface is composed of two identical $L\times L$ squares.
We have calculated the entanglement entropy for regions orthogonal to the boundaries of sizes up to $50^2\times 600$, such that $L\in [5, 50]$ and $L_y^A=600$. The lattice results are shown in \mbox{\hyperref[figortho3D]{Fig.\,\ref{figortho3D}}} where we have plotted the logarithmic contribution (divided by eight since there are eight boundary-corners $b_{3d}(\pi/2)$) against $L_{max}$. We employed the same fitting procedure presented earlier, only adapted to $d=3$ dimensions. 
We obtain from our best fit and extrapolation:
\be
b_{3d}(\pi/2) &=& 0.010416(6) \simeq \frac{1}{96}\,.\quad
\ee
We thus find excellent agreement, to the sixth significant digit, between the lattice numerics and the field theory calculation.
We intend to study further in subsequent work the angle(s) dependence of $b_{3d}(\alpha,\beta,\gamma)$, as rich universal features may be expected. 
\begin{figure}[h]
\centering
\includegraphics[scale=1]{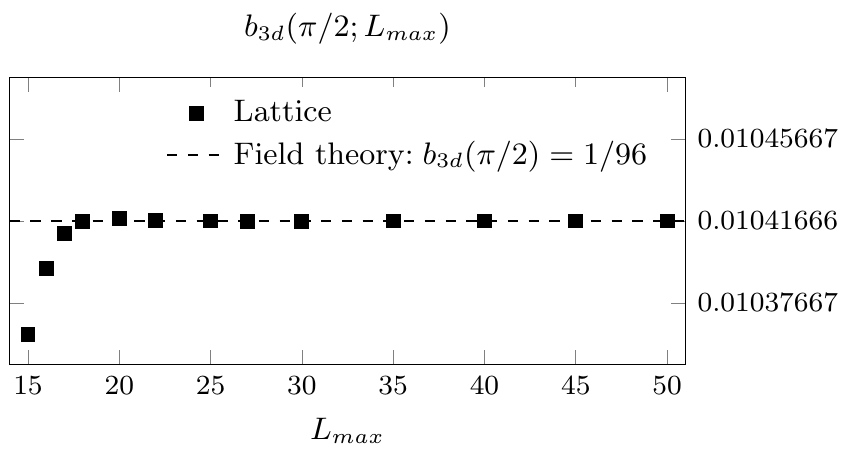}
\vspace{-15pt}
\caption{Convergence analysis of the fitted values (black squares) of $b_{3d}(\pi/2)$ for the free boson with Dirichlet BCs. PBCs are imposed in the $y-$direction (direction normal to the entangling surface) and we set $L_1=L_2=L$, and $p_{max}=4$.} 
\lb{figortho3D}
\end{figure}

\section{Conclusion}

We have presented in this paper a numerical study of the universal term in the entanglement entropy that arises due to the intersection of the entangling surface with the boundary of the space in three dimensions. This boundary-corner contribution, not to be confused with the sharp corner one due to singularities in the entangling surface, is logarithmic thus subleading to the area law, with universal coefficient $b(\theta)$. The boundary-corner function depends on the angle $\theta$ of intersection between the entangling surface and the boundary, and of course on the boundary conditions. 
Focusing on free bosons on two dimensional square lattices with Dirichlet BCs, we have performed exact numerical calculations of $b(\theta)$ for a range of angles between $\theta=0$ and $\theta=\pi/2$.
We then compared our lattice results to two candidate functions: FS formula $b_{FS}$ \eqref{sloggen} and the holographic boundary-corner function $b_\alpha$ \eqref{bholo}. For $\alpha\simeq2.56(3)$, the latter turns out to agree exceptionally well with our lattice data over the whole range of angles we considered. The (normalized) holographic boundary-corner function $b_{\alpha}(\theta)$ has an additional parameter $\alpha$, which from a mathematical point of view controls the slope of the brane in the bulk whose boundary coincides with the boundary of the BCFT$_3$, but from a BCFT perspective should be related to the boundary conditions of the underlying holographic field theory. It is therefore remarkable that for a particular value of $\alpha$, the holographic function reproduces so well the results of a free massless scalar field with Dirichlet BCs, this despite \mbox{the obvious dissimilarities of the two field theories.}

Our lattice approach allows us to probe the limiting regimes $\theta\simeq\pi/2$ and $\theta\rightarrow0$. For the orthogonal limit $b(\theta\simeq\pi/2)=a/24+\sigma(\theta-\pi/2)^2$, we obtain the numerical value $\sigma^D=0.023(4)$, while in the opposite limit $b(\theta\rightarrow0)\simeq\kappa/\theta$ we find $\kappa^D=0.044(4)$. The orthogonal regime is of particular interest. Indeed, the holographic study of the boundary-corner function \cite{Seminara:2017hhh} suggests that the coefficient $\sigma$ could be a universal quantity related to the boundary central charge $A_T$ in the one point function of stress tensor through $\sigma_E/A_{T,E}=-\pi$. We find that this ratio is violated for the free boson with Dirichlet BCs. However, one may conjecture this ratio to still be valid for free bosons with half Dirichlet and half Robin BCs due to properties that this theory shares with the holographic one. This would yield for the free boson with Robin BCs: $\sigma^R\simeq-0.0078$. 
Checking the validity of the holographic ratio in additional theories and for various BCs is an exciting issue to address. Free fermions are then next in line for consideration. We also find interesting to extend these results to the R\'enyi entropies. Furthermore, as our numerical results suggest, a complete analytical expression of $b(\theta)$ for free fields has yet to be found. Studying boundary-corners in higher dimensions is a promising path as well---we have only tackled in this paper the four dimensional case in the simple setup of orthogonality. There is still much to explore.

\begin{acknowledgments}

I wish to thank S. Solodukhin, D. Fursaev, E. Tonni, J. Sisti and D. Seminara for interesting discussions and comments on the draft. I am indebted to J. Sisti for sharing with me his Mathematica code of the holographic boundary-corner function from \cite{Seminara:2017hhh}, and to \mbox{J.-M.} St\'ephan and W. Witczak-Krempa for discussions and feedback that helped improve this manuscript.

I am grateful to the Galileo Galilei Institute for Theoretical Physics for hospitality during the program \textit{Entanglement in Quantum Systems} and the INFN for partial support along the course of this work. I also thank Perimeter Institute for the warm hospitality during the completion of this project.
This work was supported in part by the National \mbox{Natural} Science Foundation of China (NSFC, Nos. 11335012, 11325522, 11735001), and by a Boya Postdoctoral Fellowship at Peking \mbox{University.} 

\end{acknowledgments}

\appendix
\addtocontents{toc}{\protect\setcounter{tocdepth}{1}}

\section{Field theory calculations for orthogonal intersections in $d+1$ dimensions}\lb{apdxB}

\subsection{Geometrical setup}
\begin{figure}[h]
\vspace{-0.2cm}
\begin{center}
\includegraphics[scale=1]{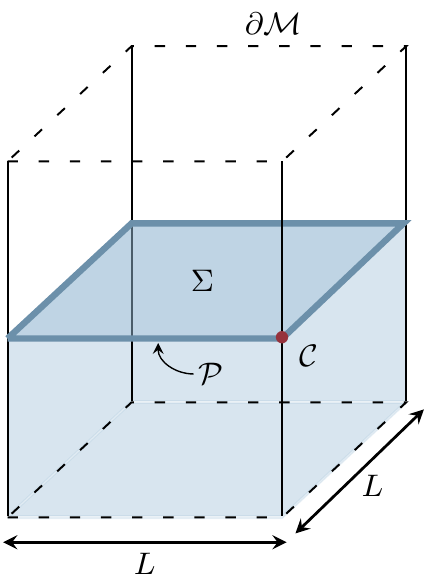}
\end{center}
\vspace{-0.6cm}
\caption{$3+1$ dimensional flat space with square cylindrical boundary $\DM$ ($t=0$ slice). The region $A$ is in light blue. The entangling surface $\Sigma$ is a plane orthogonal to the boundary $\partial\cal{M}$. Their intersection $\P$ is a rectangle (in thick blue). A boundary-corner $\mathcal{C}$ is marked by a red dot.}
\label{fig1Ap}
\end{figure}

We choose to focus on one specific geometry which captures the subleading boundary-induced terms only: the $(d+1)$-dimensional flat spacetime $\M$ has an infinite square cylindrical boundary $\DM$ and we compute the entanglement entropy for half of this infinite square cylinder (the region $A$), see \mbox{\hyperref[fig1Ap]{Fig.\,\ref{fig1Ap}}}. The distinguished region $A$ is bounded by the entangling surface, denoted $\Sigma$, which intersects orthogonally the boundary at $\P = \Sigma \cap \DM$. In 2+1 dimensions, $\Sigma$ is a line, in 3+1 dimensions it is a square, in 4+1 dimensions a cube, and so on. 

\subsection{Replica trick and heat kernel method}
A convenient way to compute the entanglement entropy in QFT is to rely on the replica trick \cite{Callan:1994py,Calabrese:2004eu,Casini:2009sr,Holzhey:1994we} and the heat kernel method. The replica trick maps R\'enyi's entropy to the partition function of the field theory on an $\a$-fold covering space $\M_\a$ with a conical singularity along the boundary $\Sigma$ of a region $A$: 
\be
S^{(\a)}(A) = \frac{1}{1-\a}\ln\tr \rho_A^\a= \frac{1}{1-\a}\ln\frac{Z(\a)}{Z^\a}\,,\quad
\lb{EER}
\ee
where $Z(\a)$ is the partition function of the theory on $\M_\a$ and $Z(1)=Z$. In the limit $\a\rightarrow 1$, the R\'enyi entropy reduces to the entanglement entropy,
\be
\lim_{\a \to 1} S^{(\a)}(A)= S(A)\,.
\lb{renyi}
\ee
At the one-loop level, the partition function $Z$ (on a manifold $\M$) can be computed via the trace of heat kernel $K$ for the Laplace-type operator $\Delta$ that describes the field theory \cite{Birrell:1982ix}, 
\be
\ln Z = \frac{1}{2}\int_{\epsilon^2}^\infty \frac{ds}{s}\, \tr\hspace{1pt}K(s)\,, \lb{schw}
\ee
where $\epsilon\rightarrow 0$ is a UV cut-off. The trace of the heat kernel encodes information about the spectrum of $\Delta$, and therefore also about the BCs.
The underlying manifold $\M$ being arbitrary, if one computes the trace of the heat kernel on the replicated spacetime $\M_\a$ which has a conical singularity along a co-dimension two hypersurface $\Sigma$, then combining (\ref{schw}) with the replica formula \eqref{EER}, one obtains the entanglement entropy associated with $\Sigma$. We denote the entanglement entropy as a function of $\Sigma$ or as a function of the region $A$, keeping in mind that it effectively depends on the geometry of the entangling surface.

\subsection{Heat equation and the method of images}

The method of images is useful to find solutions of PDE on certain domains with boundaries. The symmetries of the domain can be exploited in constructing solutions using their free space counterparts. 
Let us put this method to use to solve the heat equation on flat space with plane-parallel boundaries. 

The heat kernel $K$ satisfies the heat equation  
\be
(\partial_s - \nabla^2)K(s,x,x') = 0\,,
\ee
with the initial condition $K(s=0,x,x') = \delta(x,x')$. The solution on $\mathbb{R}$ is well-known,
\be
K_\infty(s,x,x') = \frac{1}{\sqrt{4\pi s}}e^{-\frac{1}{4s}(x-x')^2}\,.
\ee
Let us consider now the one-dimensional heat equation on $0\le x\le L$ and impose Neumann or Dirichlet boundary condition at $x=0$ and $x=L$,
\be
\partial_{x}K^{(N)}\Big|_{x=0,\,L} = 0 \,, \quad {\rm or} \quad  K^{(D)}\Big|_{x=0,\,L} = 0 \,. 
\ee
To construct the solutions corresponding to these BCs, one considers a point $P$ with position $0<x<L$ and finds its images across the two parallel planes $x=0$ and \mbox{$x=L$}. There is an infinite number of images, with positions $2Lk\pm x$, $k\in\mathbb{Z}$. For (Dirichlet) Neumann BCs, one has to (anti-) symmetrize $K_\infty$ with respect to $x=0$ and \mbox{$x=L$}, such that the solutions to the heat equation for these BCs read
\be
K^{N(D)}(s,x,x') &=& \\\nonumber
&&\hspace{-2cm}\sum_{k\in\mathbb{Z}} K_\infty(s,2Lk+x,x') \pm K_\infty(s,2Lk-x,x')\,,\quad
\ee
where the plus (minus) sign corresponds to Neumann (Dirichlet) BCs. Similarly, if we impose Neumann BCs at $x=0$ and Dirichlet at $x = L$, i.e. mixed BCs, one has
\be
K^{mixed}(s,x,x') &=& \\\nonumber
&&\hspace{-2.7cm}\sum_{k\in\mathbb{Z}} (-1)^k\Big(K_\infty(s,2Lk+x,x') - K_\infty(s,2Lk-x,x')\Big)\,,
\ee
Generalization to higher dimensions is straightforward.

\subsection{Factorization of the heat kernel and partition function}
We work in $(d+1)$-dimensional flat space with Cartesian coordinates \mbox{$X^\mu= (\tau, y, x_i, i = 1, .., d-2)$}. The entangling surface $\Sigma$ is defined by the equations $\tau = 0$ and $y=0$. The subspace $(\tau, y)$ will therefore be the two dimensional cone $C_2^{(\a)}$ with angular deficit $2\pi(1-\a)$. For a massive scalar field, the Hilbert space and the field operator $\Delta = - \nabla^2+m^2$ on the domain $\Omega = C_2^{(\a)} \times^{d-1}_{i=1}\Omega_i$ with $\Omega_i = (0,L_i)$ factorize, and so does the associated heat kernel. In this context, the method of images is particularly suitable to compute the heat kernel (see e.g. \cite{Berthiere:2016ott}). The BCs that we shall impose in each direction $x_i$ are either Neumann or Dirichlet or mixed (i.e. Neumann-Dirichlet).
The trace of the heat kernel on $\Omega$ with a conical singularity takes the compact form (we omit a volume term irrelevant to our discussion)
\be
\tr\,K_{\a}(s) &=& \frac{\a(\a^{-2}-1)}{12(4\pi)^{(d-1)/2}} \frac{e^{-sm^2}}{s^{(d-1)/2}}\\
&&\hspace{-45pt}\times \prod_{i=1}^{d-1}\left[\sum_{k_i\in\mathbb{Z}}\omega_i\int_0^{L_i}dx_i \left(e^{-\frac{L_i^2}{s}k_i^2} + \eta_i\, e^{-\frac{(x_i-L_ik_i)^2}{s}} \right)\right],\quad\nonumber
\lb{K}
\ee
where $\omega_i$ and $\eta_i$ depend on the type of BCs,
\be
{\rm Neumann/Dirichlet} &:& \quad \omega_i = 1 \,, \quad\qquad\hspace{3.8pt} \eta_i = \pm 1\,,\\
{\rm Mixed} &:& \quad \omega_i = (-1)^{k_i} \,, \quad \eta_i = - 1\,.\qquad\;\;.
\ee
%
The partition function then reads
\be
\ln Z(\a)&=&\frac{\a(\a^{-2}-1)}{24(4\pi)^{(d-1)/2}} \int_{\eps^2}^{\infty}ds \frac{e^{-sm^2}}{s^{(d+1)/2}}\nonumber\\
&&\qquad \times \prod_{i=1}^{d-1}\bigg(L_i\,\theta_{(i)}\big(e^{-L_i^2/s}\big)+\mathcal{B}_i\sqrt{\pi s}\bigg),\qquad
\lb{Zn}
\ee
where we defined
\be
\mathcal{B}_i&=&
\begin{cases}\displaystyle
\pm 1 \, , &\quad {\rm for} \;\, {\rm Neumann/Dirichlet \;BCs}\,,\quad \\
 \displaystyle \;\;0\, ,& \quad {\rm for} \;\, {\rm mixed \;BCs}\,,
\end{cases}\quad
\ee
and $\theta_{(i)}(z)$ are Jacobi theta functions such that $\theta_{(i)}= \theta_{3}$ for Neumann/Dirichlet BCs and $\theta_{(i)}= \theta_{4}$ for mixed BCs.

\subsection{Entanglement entropy}

Inserting \eqref{Zn} in \eqref{EER} and taking the limit $\a\rightarrow 1$ yields \cite{BerthierePHD}
\be
S_d(\Sigma) &=&  \frac{1}{12(4\pi)^{(d-1)/2}} \int_{\eps^2}^{\infty}ds \frac{e^{-sm^2}}{s^{(d+1)/2}}\nonumber\\
&&\qquad \times \prod_{i=1}^{d-1}\bigg(L_i\,\theta_{(i)}\big(e^{-L_i^2/s}\big)+\mathcal{B}_i\sqrt{\pi s}\bigg).\qquad
\lb{SN}
\ee
In the massless case, the entanglement entropy given in (\ref{SN}) may display IR divergences depending on the BCs. For any $d$, we find that there is no IR divergence if the BCs are not Neumann in every directions $x_i$. Thus, among the $\frac{d}{2}(d+1)$ possible BCs combinations, there is only one that leads to an IR divergence.

Integrating (\ref{SN}) over $s$ with $m=0$ and taking the limit $\eps\rightarrow 0$, the entanglement entropy for the bi-partition of an infinite square cylinder as depicted in \mbox{\hyperref[fig1Ap]{Fig.\,\ref{fig1Ap}}} reads
\be
S_{d}(\Sigma) &=& s_{d-1}\frac{\mathcal{A}_{d-1}}{\eps^{d-1}} + s_{d-2} \frac{\P_{d-2}}{\eps^{d-2}} + \cdots \lb{SNexp}\\\nonumber
&& \qquad\cdots +\,  s_{1} \frac{\P_{1}}{\eps}  -  s_{log}^{(d)}\ln\eps+ s_0 \,.\qquad
\ee
where $\mathcal{A}_{d-1}= L_1L_2\cdots L_{d-1}$ is the area of $\Sigma$. The coefficients $s_k$ are dependent on the regularization procedure thus non-universal. 
\noindent The coefficients $\P_n$ are given by \cite{BerthierePHD}
\be
\P_{d-1-p} &=& \frac{2^p}{p!(d-1-p)!}\sum_\sigma\B_{\sigma_1}\cdots\B_{\sigma_p}L_{\sigma_{p+1}}\cdots L_{\sigma_{d-1}}\,,\nonumber\\
&&
\ee
where the sum extends over all permutations of $\{ 1, \cdots, d-1\}$.
For pure Neumann or Dirichlet BCs, $\P_n$ is the $n$--area of $\Sigma$ (for $L_i=L$, \mbox{$\P_n = 2^{d-1-n}\binom{d-1}{n}L^n$}). For example, in $3+1$ dimensions, $\Sigma$ is a rectangle which has four edges (1--faces) and four corners (\mbox{0--faces}). Thus $\P_1=2(L_1+L_2)$ is the perimeter of $\Sigma$, and $\P_0=4$ its number of corners. In general, the logarithmic coefficient is proportional to the number of corners $\P_0 = 2^{d-1}$ of $\Sigma$,
\be
s_{log}^{(d)} =\,\frac{\P_{0}}{6\times2^{2(d-1)}}
\begin{cases}\displaystyle
\quad 1\, , & {\rm Neumann}\,, \vspace{8pt}\\
 \displaystyle\; (-1)^{d-1}\, ,& {\rm Dirichlet}\,, \vspace{8pt}\\
 \displaystyle\; (-1)^{n}\, ,& {\rm D^{(n)}-N^{(d-1-n)}}\,,\quad \vspace{8pt}\\
 \displaystyle\quad 0\, ,& {\rm else}\,.
 \end{cases} \quad\lb{slogd}
\ee
%
%
%
\noindent It is interesting to note that the logarithmic divergence disappears if the BCs are mixed in at least one direction. The subleading terms in \eqref{SNexp} are due to the presence of the boundary. More precisely, these contributions arise because the entangling surface intersects the boundary of the space. They are thus defined only at $\P$. 
Setting \mbox{$L_i=L$}, one may also compute for pure Neumann or Dirichlet BCs the finite part that combines with the logarithmic term (and/or the IR one),
\be
s_0 =
\begin{cases}\displaystyle
\;\frac{1-2^{d-1}}{6\times2^{d-1}}\ln L\, , &\quad {\rm Neumann}\,, \vspace{8pt}\\
 \displaystyle\; \frac{(-1)^{d-1}}{6\times2^{d-1}}\ln L\, ,& \quad {\rm Dirichlet}\,.
\end{cases} \quad
 \ee
This term is due to a combination of two factors: the intersection of the $\Sigma$ with the boundary $\partial\M$ and the BCs imposed on it, as well as the finite size of $\Sigma$.
%



%

\end{document}